\documentclass[a4paper,fleqn,usenatbib]{mnras}
\pdfoutput=1 
\usepackage{graphicx}
\usepackage{amsmath,amssymb,amstext}
\usepackage[LGR,T1]{fontenc}
\usepackage{newtxtext,newtxmath}
\usepackage[figure,figure*]{hypcap}
\usepackage{booktabs}

\graphicspath{{figures/}}
\DeclareGraphicsExtensions{.pdf}

\newcommand*{\hmpc}{h^{-1}\text{Mpc}}
\newcommand*{\hkpc}{h^{-1}\text{kpc}}
\newcommand*{\Msub}{M_\text{halo}}
\newcommand*{\Msun}{\text{M}_\odot}
\newcommand*{\Lsun}{\text{L}_\odot}
\newcommand*{\hMsun}{h^{-1}\Msun{}}
\newcommand*{\msinvsq}{\text{m}\,\text{s}^{-2}}
\newcommand*{\gobs}{g_\text{obs}}
\newcommand*{\gbar}{g_\text{bar}}
\newcommand*{\gdagger}{g_\dagger}

\newcommand*{\email}[1]{\href{mailto:#1}{#1}}
\def\equationautorefname~#1\null{Equation~(#1)\null}

\title[The RAR in MB-II]{The Radial Acceleration Relation in Disk Galaxies in the MassiveBlack-II Simulation}

\author[A.~Tenneti et al.]
{Ananth Tenneti$^{1}$\thanks{E-mail: \email{vat@andrew.cmu.edu}},
 Yao-Yuan~Mao$^{2}$\thanks{E-mail: \email{yymao.astro@gmail.com}},
 Rupert~A.~C.~Croft$^{1}$,
 Tiziana~Di~Matteo$^{1}$,
\newauthor
 Arthur~Kosowsky$^{2}$,
 Fernando~Zago$^{2}$,
and
 Andrew~R.~Zentner$^{2}$
 \\
$^{1}$McWilliams Center for Cosmology and Department of Physics, Carnegie Mellon University, Pittsburgh, PA 15213, USA \\
$^{2}$Department of Physics and Astronomy and Pittsburgh Particle Physics, Astrophysics, and Cosmology Center (PITT PACC), \\
\phantom{$^{2}$}University of Pittsburgh, Pittsburgh, PA 15260, USA
}
\pubyear{2017}

\begin{document}

\label{firstpage}
\pagerange{\pageref{firstpage}--\pageref{lastpage}}
\maketitle

\begin{abstract}

A strong correlation has been measured between the observed centripetal accelerations in galaxies and the accelerations implied by the baryonic components of galaxies. This empirical radial acceleration relation must be accounted for in any viable model of galaxy formation. We measure and compare the radial accelerations contributed by baryons and by dark matter in disk galaxies in the MassiveBlack-II hydrodynamic galaxy formation simulation. The sample of 1594 galaxies spans three orders of magnitude in luminosity and four in surface brightness, comparable to the observed sample from the Spitzer Photometry \& Accurate Rotation Curves (SPARC) dataset used by \citet{2016arXiv160905917M}. We find that radial accelerations contributed by baryonic matter only and by total matter are highly correlated, with only small scatter around their mean or median relation, despite the wide ranges of galaxy luminosity and surface brightness. We further find that the radial acceleration relation in this simulation differs from that of the SPARC sample, and can be described by a simple power law in the acceleration range we are probing.

\end{abstract}


\section{Introduction}

In the current standard cosmological model ($\Lambda$CDM), nearly 85\% of the Universe's matter content is composed of dark matter. This picture has been consistently corroborated by a variety of cosmological probes, such as the cosmic microwave background \citep{ade2014planck}, the growth of large-scale structure \citep{blumenthal1984formation}, and Big Bang nucleosynthesis yields for light elements \citep{walker1991primordial}. On galaxy and cluster scales, gravitational lensing provides further evidence for the existence of dark matter with roughly the same relative abundance as that inferred from cosmology \citep{2013MNRAS.430.2200K,2016PhRvD..94b2001A,2017MNRAS.465.2033J}.

Early evidence of dark matter was found in the context of galactic dynamics. \citet{zwicky1933rotverschiebung} suggested its existence as an explanation for the discrepancy between the luminous matter and the dynamical mass in the Coma cluster. 
A case for dark matter via measurements of galaxy rotation curves was made later by \citet{rubin1970rotation}. Rotation curves with the same characteristic shape have now been observed for thousands of spiral galaxies \citep[e.g.,][]{persic1996universal}.

Recently, \citet{2016arXiv160905917M} employed the new Spitzer Photometry \& Accurate Rotation Curves (SPARC) database \citep{lelli2016sparc} to report a correlation between the radial acceleration traced by rotation curves ($\gobs$) and the radial acceleration inferred from the observed distribution of baryons ($\gbar$). 
This correlation has long been known \citep[see e.g.,][]{1990A&ARv...2....1S,2000ApJ...533L..99M}; however, the new analysis is particularly revealing because previous uncertainties in baryonic mass-to-light ratios are mitigated using infrared imaging. The SPARC analysis exploited a sample of 153 galaxies with a wide range of luminosities, effective surface brightnesses, and gas fractions. The authors found that the Radial Acceleration Relation (RAR) displayed by the data 
is well described by
\begin{equation} \label{eq:gobsgbar}
\gobs = \frac{\gbar}{1 - \exp \left(-\sqrt{ \gbar/\gdagger } \right)},
\end{equation}
where $\gdagger = 1.20 \pm 0.26 \times 10^{-10}\, \msinvsq$ corresponds to the acceleration scale below which $\gobs$ deviates from $\gbar$, with little scatter ($0.11$~dex). Notably, the correlation persists when dark matter dominates 
$\gobs$. This implies that the distribution of dark matter is completely determined by that of baryons or vice versa.

The RAR over a wide range of galaxy masses and surface brightnesses is not an obvious result of standard galaxy formation scenarios in dark matter haloes. Though it may not be difficult to imagine that a relation of this kind is obtained in standard $\Lambda$CDM cosmological models \citep[e.g.,][]{kaplinghat_turner2002,navarro_etal2016}, the tightness of the observed RAR may be more difficult to understand \citep[e.g.,][]{2017MNRAS.464.4160D}. In any case, the observed RAR has led to speculation that its origin lies in an underlying departure from Newtonian gravity 
\citep[e.g.,][]{1983ApJ...270..365M,1983ApJ...270..371M}. 

Of course, it is certainly possible that the RAR is a natural outcome of galaxy formation in the standard cosmological model and any claims to the contrary should be based upon specific predictions of the standard cosmological model. Only recently have simulations including dark matter, hydrodynamics, star formation, and various forms of baryonic feedback attained sufficient mass resolution to form realistic disk galaxies \citep[see e.g.,][]{2014Natur.509..177V,schaye2015eagle,2015MNRAS.450.1349K} such that the RAR can be studied within the context of the $\Lambda$CDM model for structure formation. 

Indeed, \citet{2016arXiv161006183K} and \citet{2016arXiv161007663L} studied the RAR in the MUGS2 \citep{stinson2010cosmological,keller2014superbubble,keller2015cosmological} and EAGLE \citep{schaye2015eagle} simulations, respectively. While \citet{2016arXiv161006183K} were able to recover the RAR found in \citet{2016arXiv160905917M}, their sample of 18 galaxies with masses ranging from $1.8 \times 10^{10}$ to $2.7 \times 10^{11} \, \Msun$ does not include representatives of the fainter galaxies present in SPARC. Moreover, the MUGS2 sample is sufficiently small that deviations from and/or scatter about the RAR could be missed simply due to finite sampling.  

\citet{2016arXiv161007663L} obtained a tight RAR in the EAGLE simulation, with an optimal fitting parameter $\gdagger = 3.0 \times 10^{-10} \, \msinvsq$, which is 2.5 times larger than that of the observed SPARC sample. However, \citet{2016arXiv161007663L} do not quote the range of galaxy surface brightness of their simulated galaxy sample, and do not select disk galaxies, therefore it is difficult to compare the EAGLE sample with the SPARC sample in a robust manner. 

The authors of the MUGS2 and EAGLE studies reach different conclusions regarding the evolution of the RAR with redshift. 
While \citet{2016arXiv161006183K} find the RAR to evolve with redshift in the MUGS2 simulation, \citet{2016arXiv161007663L} report that the RAR remains unchanged at earlier times in the EAGLE simulation.

In this study, we use the MassiveBlack-II (MB-II) hydrodynamic simulation to analyze the RAR for disk galaxies spanning a wide range of masses and surface brightnesses, roughly comparable to those present in the SPARC data.  We calculate the radial accelerations contributed by baryonic matter only ($\gbar$) and by all matter ($\gobs$), and  find them tightly correlated over the full range of disk galaxies, with only small scatter of around $0.1$~dex, despite the wide range of luminosities and surface brightnesses spanned by the simulated galaxy sample. However, the RAR in this simulation differs from that of the SPARC sample. In particular, we find that the RAR in MB-II can be described by a single power law for accelerations 
$\gbar \lesssim 10^{-9}\, \msinvsq$. This power law is steeper 
than the low-acceleration asymptote 
$\gobs \rightarrow \sqrt{\gdagger \, \gbar}$ 
of \autoref{eq:gobsgbar}. Moreover, the SPARC data prefer a clear 
break from this asymptotic behavior at an acceleration of 
$\gdagger \simeq 10^{-10} \, \msinvsq$.

This paper is organized as follows. We introduce the MB-II simulation, the galaxy sample used in this work, and the measurement of the RAR in 
\autoref{sec:methods}. We present our results in \autoref{sec:results}, and further discuss and summarize these results in \autoref{sec:discussion}.

\section{Methods}
\label{sec:methods}

\subsection{MassiveBlack-II Simulation} \label{sec:sims}

In this study, we inspect the RAR in the MassiveBlack-II (MB-II) hydrodynamic 
simulation. MB-II is a state-of-the-art high-resolution, large-volume, cosmological hydrodynamic simulation of structure formation.
This simulation has been performed with 
{\sc p-gadget}, which is a hybrid version of the parallel code, {\sc gadget2} \citep{2005MNRAS.361..776S}, upgraded to run on Peta-flop scale supercomputers. In addition to gravity and smoothed-particle
hydrodynamics, 
the {\sc p-gadget} code also includes the physics of multiphase interstellar medium model with star formation \citep{2003MNRAS.339..289S}, black hole accretion, and AGN feedback
\citep{2005MNRAS.361..776S,2012ApJ...745L..29D}. Radiative cooling and
heating processes are also included \citep[as in][]{1996ApJS..105...19K},
as is photoheating due to an imposed ionizing ultraviolet background. The
details of this simulation can be found in \cite{2015MNRAS.450.1349K}.

MB-II contains $N_\text{part} = 2\times 1792^{3}$ dark matter and gas
particles in a cubic periodic box with a side length of $100 \, \hmpc$,
and a (Plummer-equivalent) gravitational softening length $\epsilon = 1.85 \, \hkpc$ in
comoving units. A single dark matter particle has a mass $m_\text{DM} =
1.1\times 10^{7} \, \hMsun$ and the initial mass of a gas particle is
$m_\text{gas} = 2.2\times 10^{6} \, \hMsun$, with the mass of each star
particle being $m_\text{star} = 1.1\times 10^{6} \, \hMsun$. The cosmological
parameters used in the simulation are as follows: amplitude of matter
fluctuations $\sigma_{8} = 0.816$, spectral index $n_s = 0.96$,
mass density parameter $\Omega_{m} = 0.275$, cosmological constant
density parameter $\Omega_{\Lambda} = 0.725$, baryon density parameter
$\Omega_{b} = 0.046$, and Hubble parameter $h = 0.702$ as per WMAP7
\citep{2011ApJS..192...18K}.

\subsection{Galaxy samples}
\label{sec:samples}
\begin{figure*}
\centering
\includegraphics[width=0.48\textwidth]{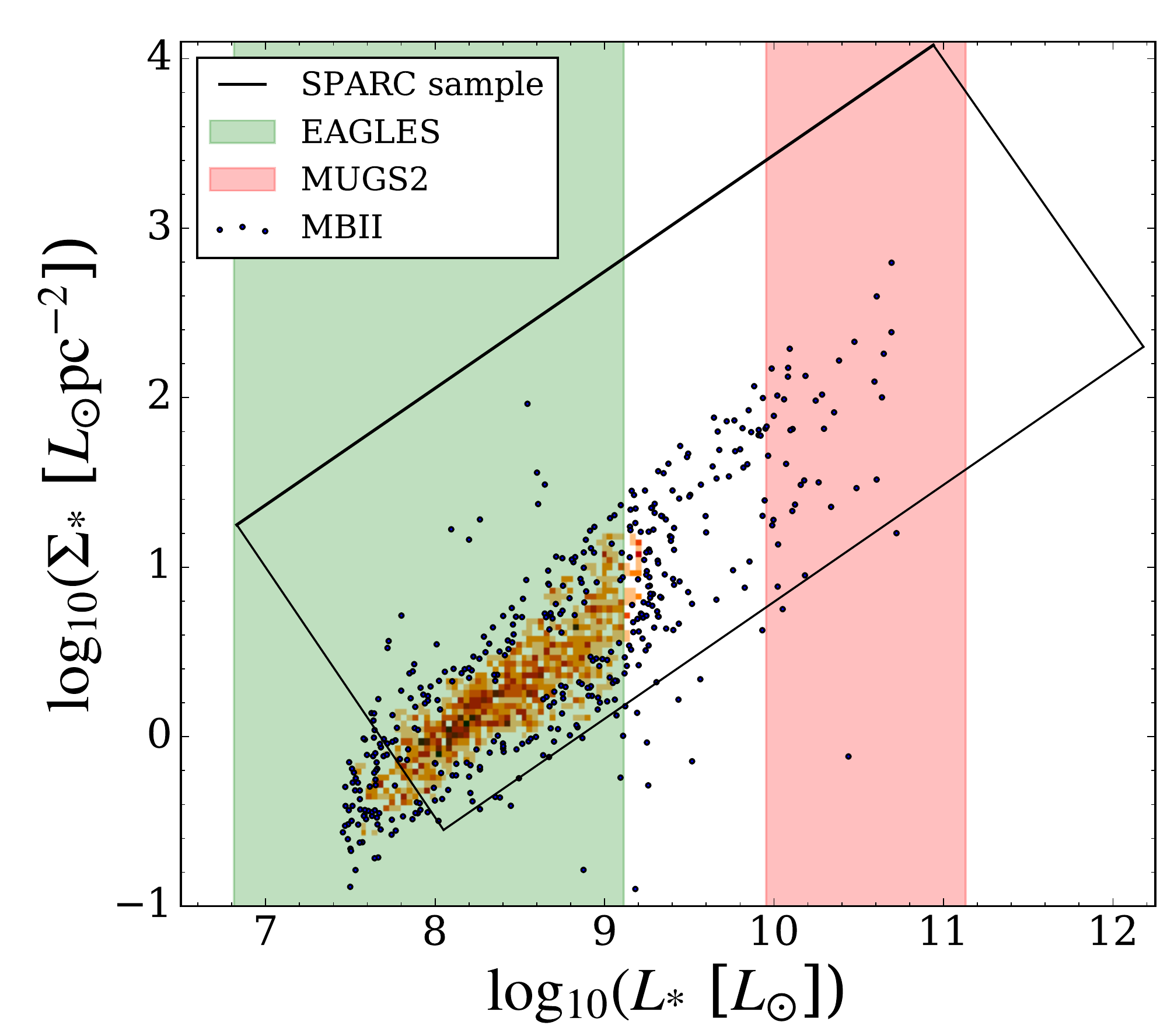} \hspace{0.03\textwidth}
\includegraphics[width=0.48\textwidth]{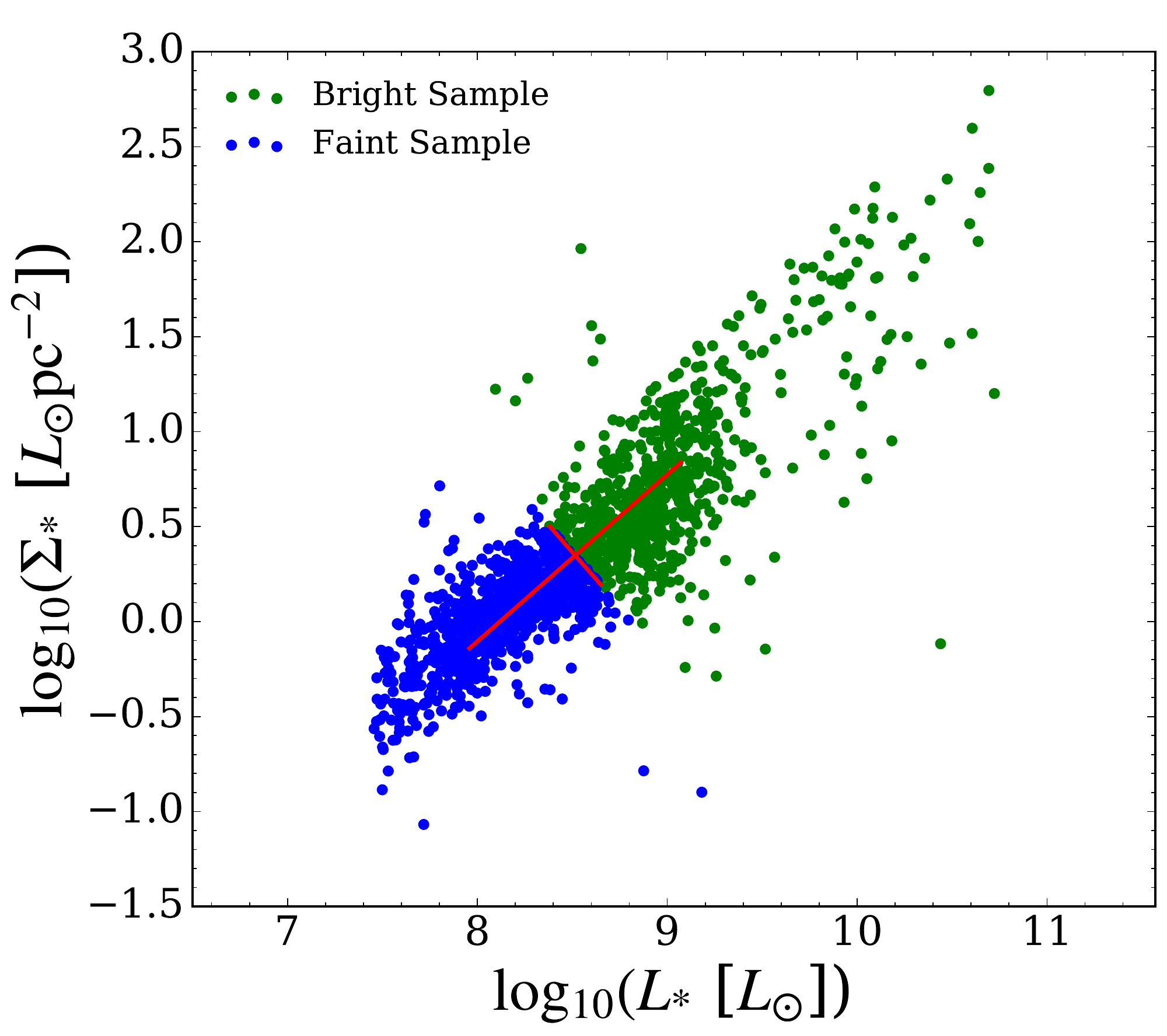}
\caption{\label{F:fig_lum}{\em Left:}  Distribution of the surface brightness of disk galaxies in the MB-II simulation as a function of galaxy luminosity. The high density regions are shown by a 2D histogram, while at lower density, the scatter depicts each galaxy. The parallelogram shows the region of distribution of SPARC galaxy sample in the $L_* - \Sigma_*$ plane. Also shown are the luminosity ranges of the simulated galaxies in EAGLE and MUGS2 samples, in the vertical green and pink bands. Note that these two bands extend the full vertical range because there is no information about the surface brightness of these two samples. {\em Right:} Scatter plot showing the split of galaxy sample in MB-II into bright and faint galaxies. The green region depicts the bright sample, while the blue region indicates the faint sample. The red lines show the principal components of the galaxy distribution.}
\end{figure*}

To identify the morphological type of a galaxy in the MB-II simulation, we follow the procedure from \cite{2009MNRAS.396..696S} and define a circularity parameter, $\epsilon \equiv j_{z}/j_\text{circ}(r)$, for each star within 10 times the stellar half-mass radius. Here $j_{z}$ is the component of the specific angular momentum of the star in the direction of the total angular momentum of the galaxy, and $j_\text{circ}(r)$ is the specific angular momentum of a circular orbit at the same radius as the star,
\begin{equation} \label{eq:jcirc}
 j_\text{circ}(r) = r \, V_\text{circ}(r) = \sqrt{GM(<r)r}.
\end{equation}
The total angular momentum of the galaxy is calculated using all star particles within 10 times the stellar half-mass radius.

All stars with $\epsilon > 0.7$ are then identified as disk stars. The bulge-to-total ratio is hence defined as $\text{BTR} = 1 - f_{\epsilon >0.7}$, where $f_{\epsilon >0.7}$ is the fraction of disk stars within 10 times the stellar half-mass radius. We then classify galaxies with $\text{BTR} < 0.7$ as disk galaxies.

In the left panel of \autoref{F:fig_lum}, we show the distribution of luminosity and surface brightness of all disk galaxies in this simulation. The luminosity is in the Spitzer $3.6 \, \mu\text{m}$ band. To compare our sample with the observed SPARC sample and other simulated samples, we also outline the range of luminosities ($10^{7} < L_{[3.6]} < 5 \times 10^{11} L_{\odot}$) and surface brightnesses ($5< \Sigma_* < 3\times 10^{3} \, \text{pc}^{-2}\,\Lsun $) of the SPARC sample with the black parallelogram, and show the range of luminosities in the MUGS2 \citep{2016arXiv161006183K} and the EAGLE \citep{2016arXiv161007663L} samples in red and green bands respectively. Note that the stellar masses are converted into luminosities based on the 
mass-to-light ratio of $0.5$ in the SPARC sample. 

Among the samples shown in \autoref{F:fig_lum}, the SPARC sample has the largest dynamical range. Our galaxy sample in MB-II spans a much larger range compared to the other two simulated samples, and is broadly comparable with the SPARC sample. Compared with the SPARC sample, the MB-II sample has fewer high luminosity and high surface brightness galaxies. This difference is likely due to both selection effects and the limited volume of the MB-II simulation.
We also note that in the MB-II sample, there seems to be a sharp decrease in population above $2 \times 10^9 \, \Lsun$. This is caused by our disk selection criteria. 
We also divide the disk sample into bright and faint galaxies along the first principal component (axis with greatest variance) in the $\log L_* - \log \Sigma_*$ plane, as shown in the right panel of \autoref{F:fig_lum}, to study the 
dependence of the RAR on luminosity.

\subsection{Measuring and fitting of the RAR}\label{sec:rar}

For each galaxy, we calculate the radial acceleration at locations on the disk plane that are logarithmically spaced in distance from the center of the galaxy. The range of distances spans $0.25 - 4.0$ times the stellar half-light radius. At each of these locations, we estimate $\gobs$ \citep[in the notation of][]{2016arXiv160905917M}, which denotes the ``observed'' radial component of the gravitational acceleration contributed by all particles in the galaxy. The quantity 
$\gobs$ includes accelerations induced by dark matter as well as those induced by gas, stars, and black holes.  
At each of these locations, we also compute $\gbar$, which denotes the gravitational acceleration contributed by all particles except for dark matter particles. We calculate both $\gobs$ and $\gbar$ via direct summation of the gravitational accelerations projected onto the radial direction in the plane of the galaxy disk. We have also confirmed that, 
for all the disk galaxies in this sample, more than 85\% of the particles have a radial acceleration that is larger than 90\% of the magnitude of its total acceleration.

While $\gobs$ and $\gbar$ as defined here share the same notations and physical motivations as in those defined in \citet{2016arXiv160905917M}, the operational definitions are distinct.
\citet{2016arXiv160905917M} determine $\gobs$ by examining rotation speeds, and $\gbar$ by estimating the gravitational potential induced by the observed baryons. By contrast, the velocities of star particles never enter our calculation of $\gobs$ since accelerations can be determined by direct summation and do not need to be inferred from measured velocities. It would be interesting to examine, in a high-resolution zoom-in simulation, how well one can recover the actual mass distribution from the observed velocity dispersion. In the present paper, we focus our study on our idealized measurements of $\gobs$ and $\gbar$.

For both of our high- and low-luminosity samples, we fit the measured RAR with two different functional forms and inspect the residuals. The first function to which we fit our data is given in \autoref{eq:gobsgbar} and is the relation used in \citet{2016arXiv160905917M}. This form has only one free parameter, $\gdagger$.
The other functional form we use here is a power-law relation,
\begin{equation} \label{eq:powerlaw}
\log_{10} \,\left(\frac{\gobs}{\msinvsq}\right) = A \log_{10} \, \left(\frac{\gbar}{\msinvsq}\right) + B,
\end{equation}
with two free parameters: the log--log slope $A$, and the intercept $B$.
While there is no physical motivation to use a power-law relation, we find that the power-law relation gives good fits to the RAR (over a restricted range of accelerations) as we will demonstrate in the next section. However, we note that the power law cannot describe the relation at large radial acceleration as we expect  $\gobs \geq \gbar$ to always hold, and hence the power law must break down at large acceleration.

Before proceeding to our primary results, it is worth noting that the fits can be affected significantly by the distribution of galaxy properties. The galaxies in the 
SPARC sample are (approximately) uniformly distributed within the parallelogram in \autoref{F:fig_lum}, whereas the 
galaxies in our simulated samples are volume limited and thus distributed quite differently in the $L_* - \Sigma_*$ plane. We weight the galaxies in our samples so as to mimic the weighting in the SPARC sample. To be specific, we identify the principal components of the galaxy distribution in the $\log L_* - \log \Sigma_*$ plane, based on a linear transformation that defines the axis with greatest variance as the first principal axis. The principal components are shown as red lines in the right panel of \autoref{F:fig_lum}. We then bin the galaxies in our sample according to the first principal component, and then 
weight the data from each galaxy by $1/N_i$, 
where $N_i$ is the population in the corresponding bin. While this improves the comparison between the MB-II and SPARC samples, it certainly does not account for the fact that the high-luminosity and high-surface brightness regions of the SPARC sample are not present in the simulated MB-II sample.

\section{Results}
\label{sec:results}

\begin{figure}
\centering
\includegraphics[width=\columnwidth,clip,trim=0.0in 0 0.0in 0]{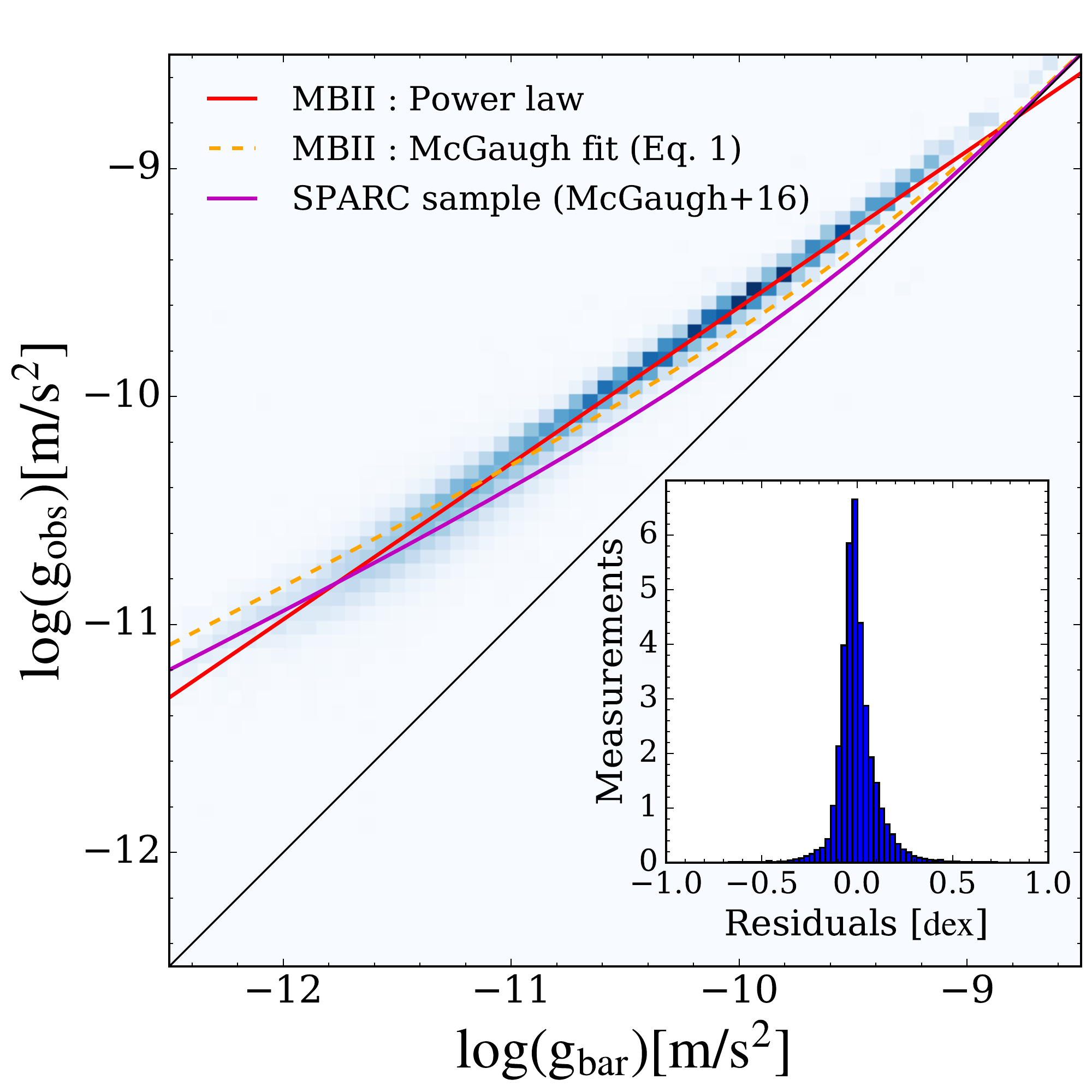} \\ 
\includegraphics[width=\columnwidth,clip,trim=0.0in 0 0in 0]{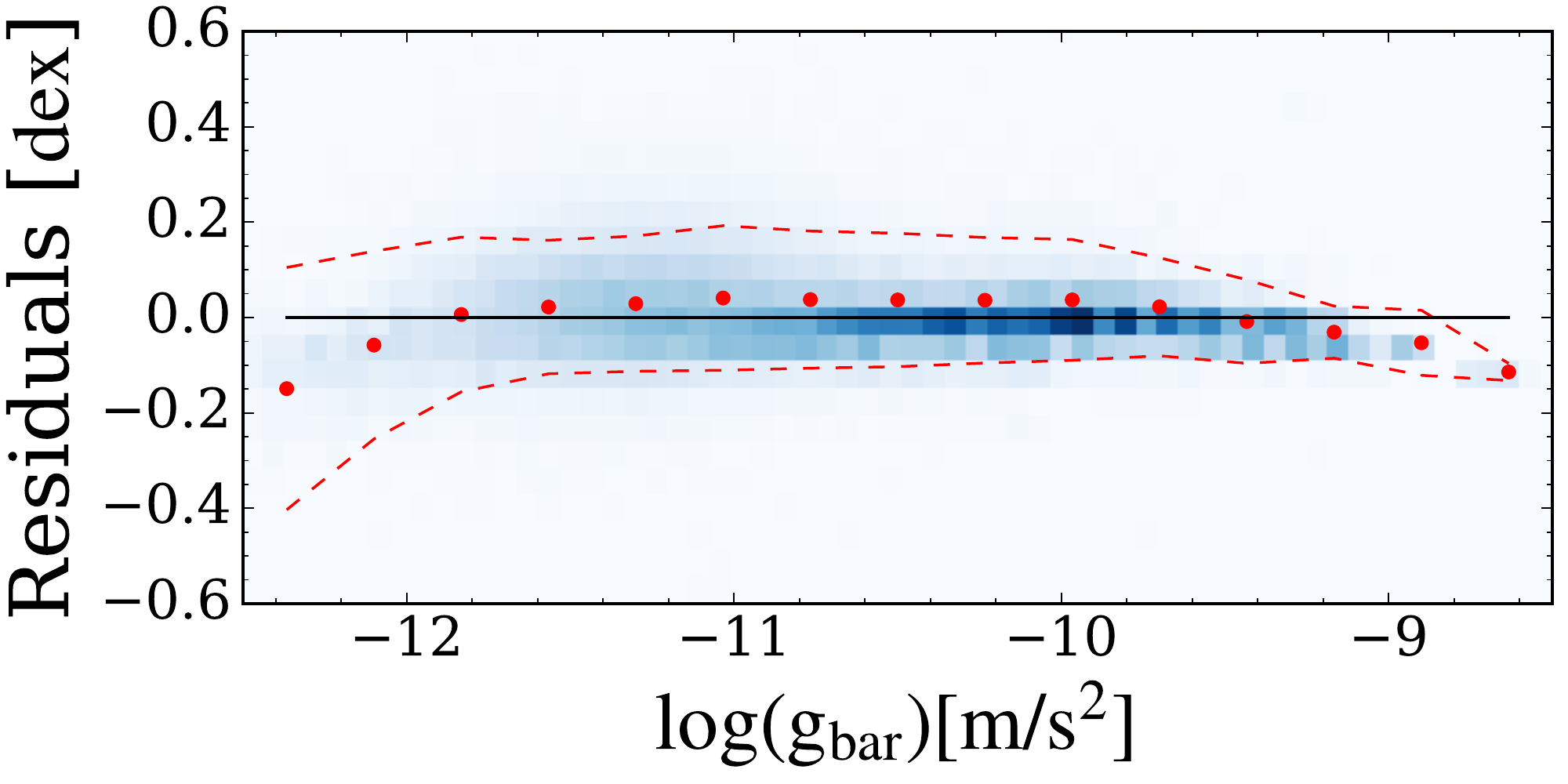}
\caption{\label{F:fig_gobsgbar_all} RAR of the baryonic component ($\gbar$) compared with that of total matter ($\gobs$) . {\em Top:} 2D weighted histogram of $(\gobs, \gbar)$ for all disk galaxies in MB-II. The red solid line shows a power law fitted to the data, and the dashed orange curve shows \autoref{eq:gobsgbar} with best-fit $\gdagger$. The observational result from SPARC sample is shown by the magenta solid line \citep{2016arXiv160905917M}. The inset shows the total weighted distribution of residual values with respect to the best-fit power law $(\gobs^\text{(fit)} - \gobs)$. 
 {\em Bottom:} The residual values with respect to the best-fit power law as a function of $\gbar$. The red points and dashed lines show the mean and $1\sigma$ of the residual distribution in each of the $\gbar$ bins.}
\end{figure}

\begin{figure*}
\centering
\begin{minipage}{.49\textwidth}
\centering
\includegraphics[width=\columnwidth,clip,trim=0.0in 0 0.0in 0]{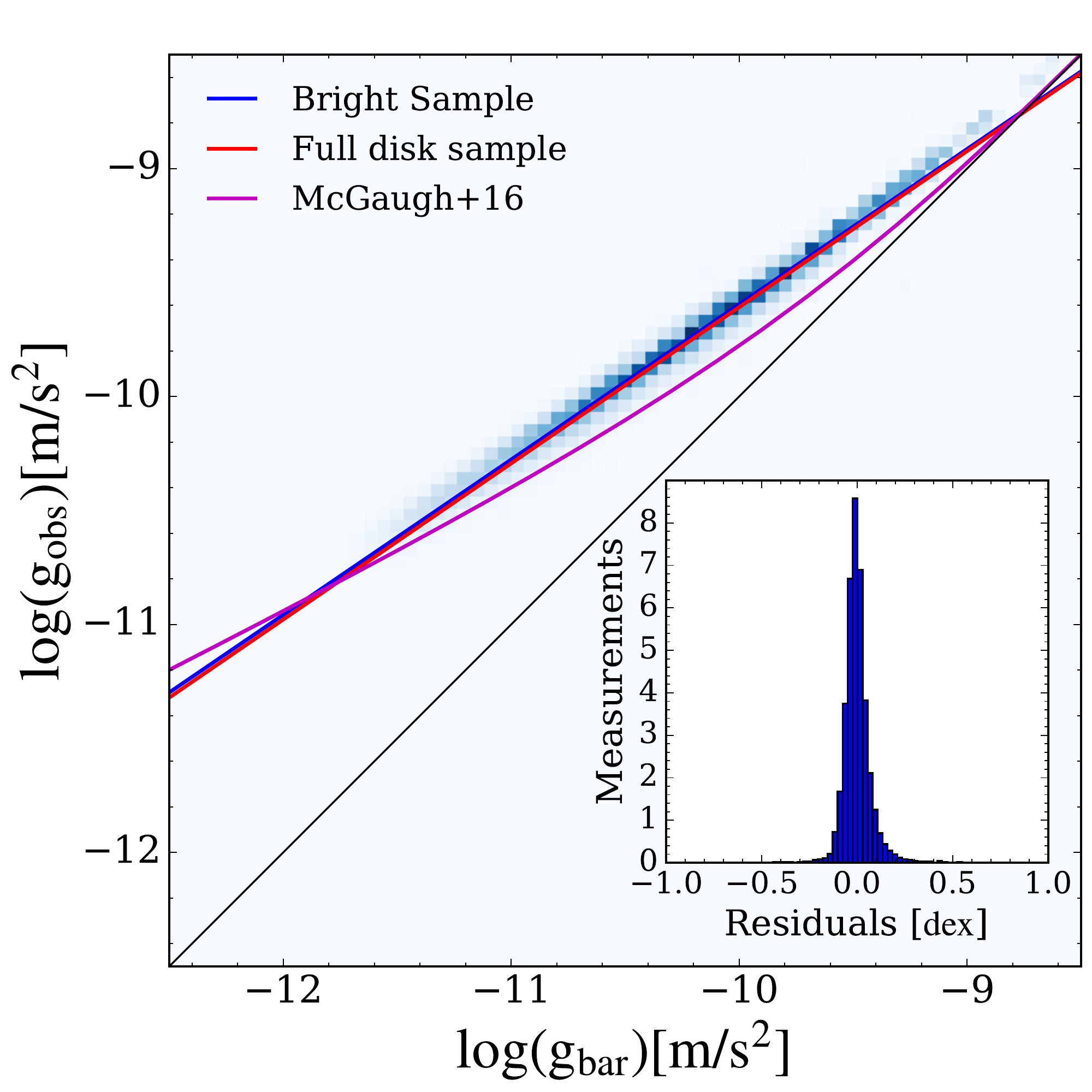} \\
\includegraphics[width=\columnwidth]{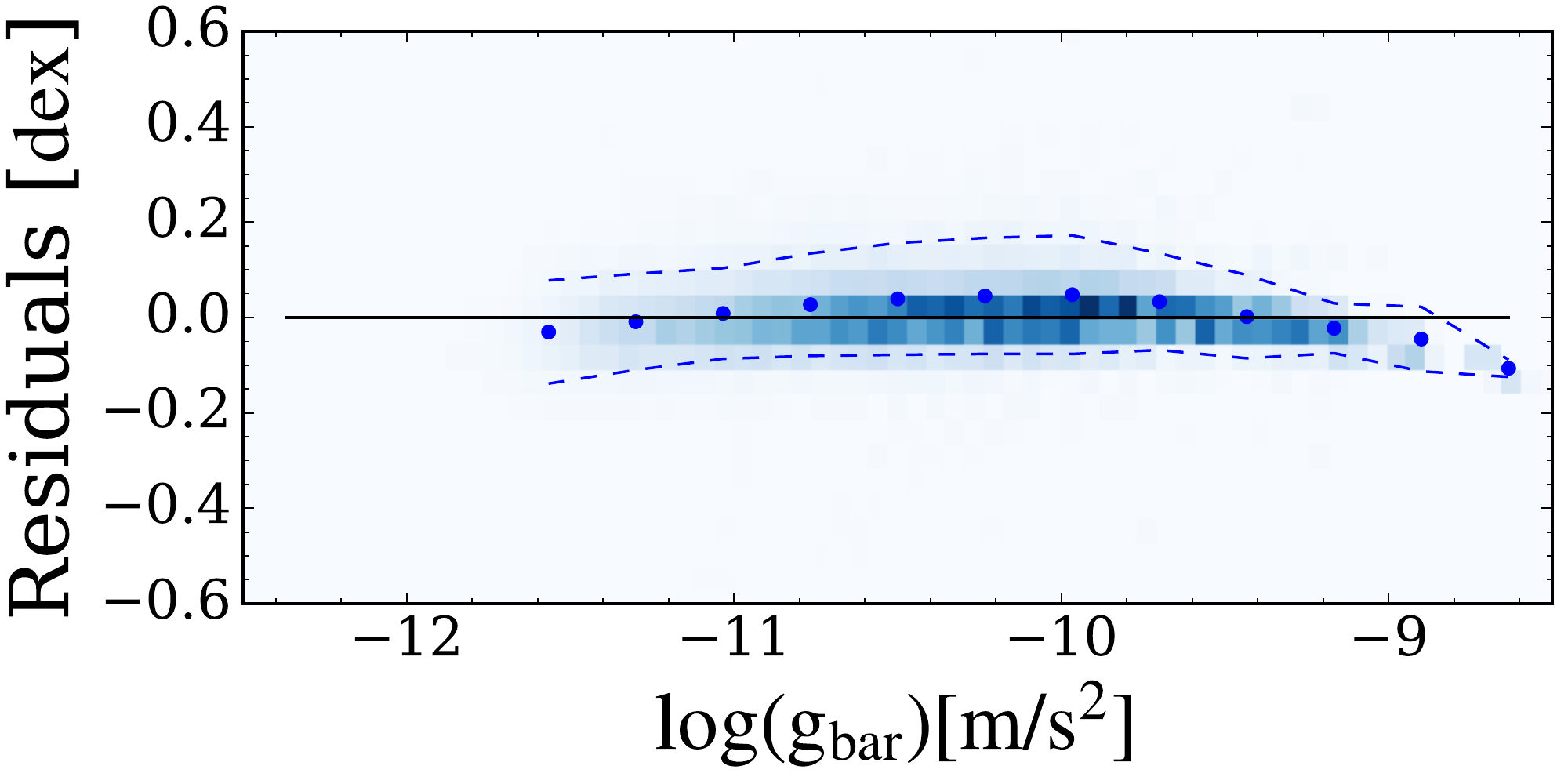}
\end{minipage}%
\begin{minipage}{.49\textwidth}
\centering
\includegraphics[width=\columnwidth,clip,trim=0.0in 0 0.0in 0]{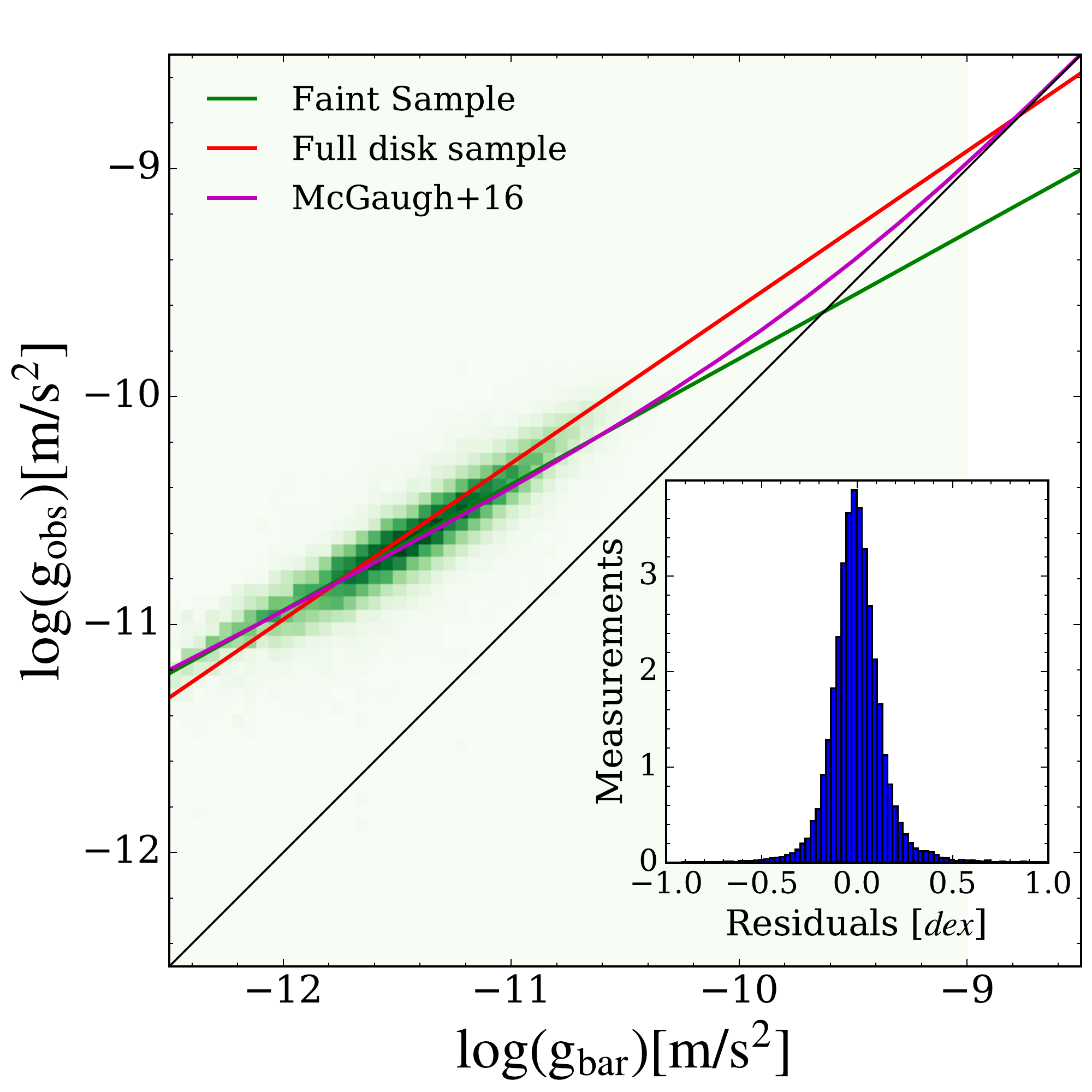} \\
\includegraphics[width=\columnwidth]{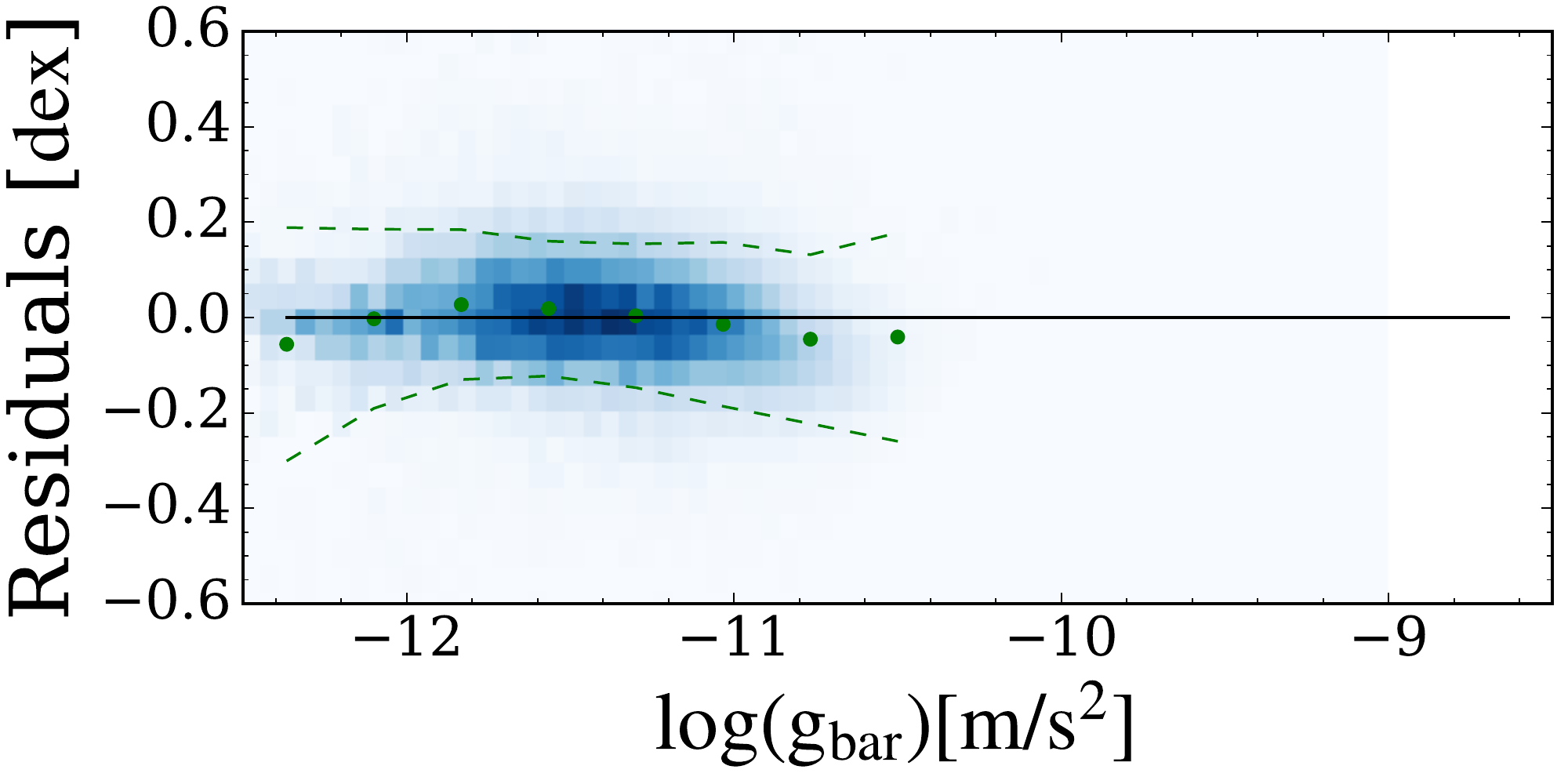}
\end{minipage}
\caption{\label{F:fig_gobsgbar_surfb} Similar to \autoref{F:fig_gobsgbar_all} but for the bright (left) and faint (right) samples of galaxies. The blue solid line in the left panel and the green solid line in the right panel show the best-fit power laws for each case respectively.
The red solid lines show the best-fit power law for the full disk sample. The magenta solid lines show the observational result from SPARC sample \citep{2016arXiv160905917M}.
The total distribution of residuals $(\gobs^\text{(fit)} - \gobs)$ is shown in the inset, and the residual as a function of $\gbar$ is shown in the bottom panels (with points and dashed lines showing the mean and $1\sigma$ of the residual distribution in each of the $\gbar$ bins).} 
\end{figure*}

\begin{figure}
\centering
\includegraphics[width=\columnwidth,clip,trim=0.0in 0 0.0in 0]{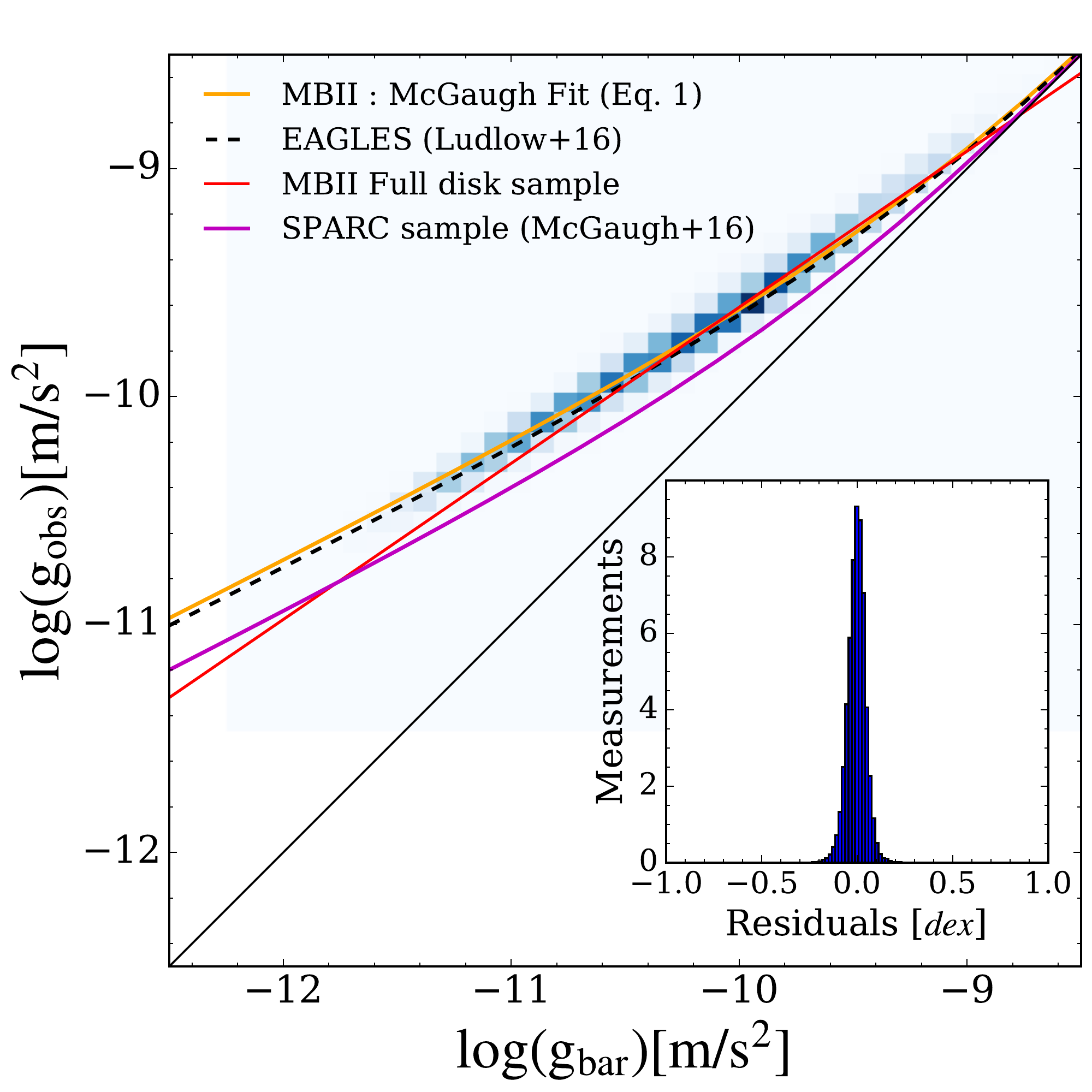} \\
\includegraphics[width=\columnwidth]{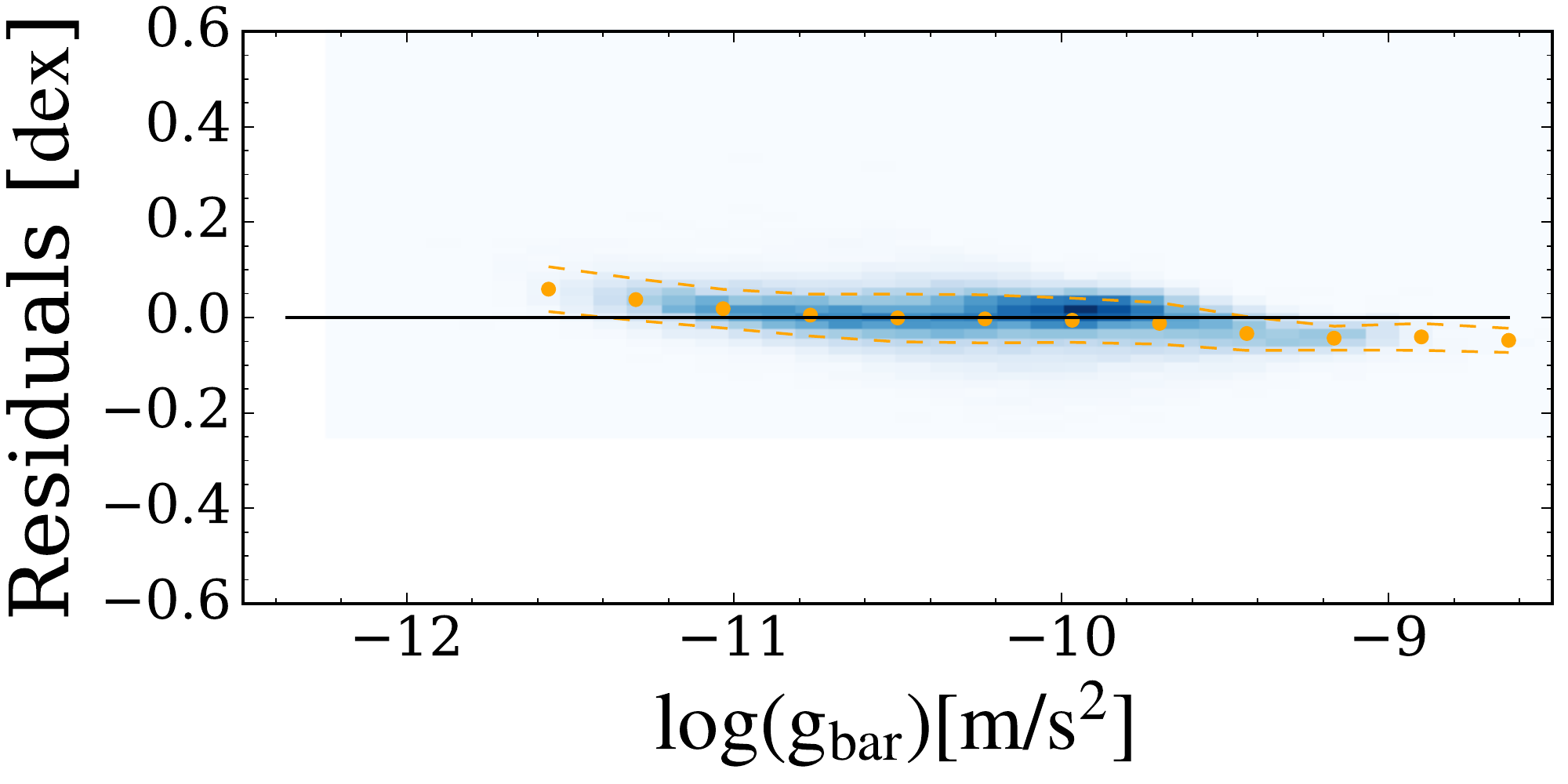}
\caption{\label{F:fig_gobsgbar_strm}Similar to \autoref{F:fig_gobsgbar_all} but for central galaxies with $\Msub > 10^{12}\, \hMsun $ to mimic the EAGLE sample. 
The solid orange line show the best fit to  \autoref{eq:gobsgbar}, and the black dashed line shows the fit obtained by \citet{2016arXiv161007663L} with the EAGLE simulation.
The total distribution of residuals $(\gobs^\text{(fit)} - \gobs)$  is shown in the inset, and the residual as a function of $\gbar$ is shown in the bottom panels (with points and dashed lines showing the mean and $1\sigma$ of the residual distribution in each of the $\gbar$ bins). }
\end{figure}

\begin{table*}
\begin{center}
\caption{\label{T:tab1} Best-fit values for $(A, B)$ in \autoref{eq:powerlaw} and for $\gdagger$ in \autoref{eq:gobsgbar}, and the corresponding residuals, for the full disk galaxy sample, disk samples split into bright and faint galaxies, and also central galaxies with subhalo mass, $\Msub > 10^{12} \, \hMsun$. The best-fit value of $\gdagger$ and residual of SPARC sample from \citep{2016arXiv160905917M} is also listed for comparison.}
\begin{tabular}{lcccccc}
\toprule
& \multicolumn{3}{c}{Power-law fit (Eq.~\ref{eq:powerlaw})} & \multicolumn{2}{c}{\citeauthor{2016arXiv160905917M} fit (Eq.~\ref{eq:gobsgbar})} & \\
\cmidrule(lr){2-4}\cmidrule(lr){5-6}
Galaxy sample & $A$ (slope) & $B$ (intercept) & Residual & $\gdagger$ [$10^{-10} \, \msinvsq$] & Residual & \# of data points\\ 
\midrule
 Full Disk Sample & $0.6849 \pm 0.0004$ & $-2.759 \pm 0.004$ & $0.013^{+0.061}_{-0.091}$ &  $2.01 \pm 0.004$ & $0.053^{+0.068}_{-0.17}$ & 159400\\
 Bright Sample &   $0.6811 \pm 0.0006$ & $-2.784 \pm 0.006$ & $0.006^{+0.047}_{-0.056}$ &  $3.05 \pm 0.007$ & $0.023^{+0.052}_{-0.084}$ & 71600\\
 Faint Sample  &  $0.5523 \pm 0.0012$ & $-4.311 \pm 0.014$ & $0.007^{+0.11}_{-0.12}$ &  $1.151 \pm 0.003$ & $0.007^{+0.11}_{-0.12}$ & 87800\\
 Centrals \footnotesize{($\Msub > 10^{12} \, \hMsun$)} &  $0.6374 \pm 0.0005$ & $-3.215 \pm 0.005$ & $0.0008^{+0.050}_{-0.049}$ & $3.491\pm 0.005$ & $0.0008^{+0.047}_{-0.041}$ & 32648\\
 SPARC sample &  - & - & - & $1.2\pm 0.02$ & $0.^{+0.055}_{-0.055}$ & 2693\\
\bottomrule
\end{tabular}
\end{center}
\end{table*}

\autoref{F:fig_gobsgbar_all} shows the RAR for all the disk galaxies in the MB-II simulation. 
We fit the $\gobs-\gbar$ relation with the two functional forms introduced in \autoref{sec:rar}, and compare them with the relation for the SPARC sample. The RAR of the simulated galaxies clearly deviates 
from that of the SPARC sample. In particular, the RAR of the simulated 
galaxies is a significantly steeper at low acceleration than that of the SPARC sample. 
As a consequence, \autoref{eq:gobsgbar} is not a good description of the 
simulated data because it asymptotes to a power law $\gobs \propto \sqrt{\gbar}$ for $\gbar \ll \gdagger$. The MB-II disk galaxies obey a relation closer to 
$\gbar \propto \gbar^{0.7}$. In fact, over the range of accelerations represented in the simulations, the RAR of the simulated disk galaxies is much better described by a single power-law relation than by \autoref{eq:gobsgbar}. At most, there is only a slight hint of a break from the power-law near an acceleration of 
$\gbar = 10^{-9} \, \msinvsq$, an order of 
magnitude larger than the transition acceleration 
$\gdagger \simeq 10^{-10}\, \msinvsq$ of the SPARC sample.
The scatter in the simulated sample is relatively small around the best-fit power-law relation, with a width in the residual distribution of only around 0.1~dex. In detail, this scatter is a function of acceleration and is 
larger for lower accelerations. The residual distribution is also slightly skewed, as is evident in the inset in \autoref{F:fig_gobsgbar_all}.

To further study how the RAR depends on luminosity and surface brightness, we also divide the disk sample into bright and faint galaxies along the first principal component in the $L_* - \Sigma_*$ plane, as shown in the right panel of \autoref{F:fig_lum}.
\autoref{F:fig_gobsgbar_surfb} shows the RARs for both samples. 
The scatter in the bright sample is much smaller than 
that of the full sample, and the median relation (fitted by a power law) of the bright sample is similar to that of the full simulated disk sample and distinct from 
the SPARC sample. The RAR of the bright sample is what drives the fits to be formally distinct from the fit to the SPARC sample, because the RAR has a much steeper slope than the asymptotic low-acceleration slope of the SPARC sample.

The faint sample, shown in the right panel of \autoref{F:fig_lum}, 
spans a significantly narrower range of accelerations.  However, over the range of accelerations where the faint simulated galaxy sample overlaps with the SPARC data, the two samples exhibit similar RARs. In particular, 
the slope of the RAR of faint galaxies in MB-II is much closer to that of 
the SPARC sample, with approximately $\gobs \propto \sqrt{\gbar}$. This is evident 
through the proximity of the green and purple curves in the right panel of \autoref{F:fig_lum}. 
The faint sample also exhibits much larger scatter in the RAR, which is likely due to the effect of finite resolution.

For a direct comparison with the EAGLE result presented in \citet{2016arXiv161007663L}, we analyze a sample of central galaxies in MB-II with a halo mass $\Msub > 10^{12} \, \hMsun$, shown in \autoref{F:fig_gobsgbar_strm}. The method to calculate radial acceleration also matches the calculation in \cite{2016arXiv161007663L}, where the $\gobs$ and $\gbar$ accelerations at a given radial distance from the center are obtained from the total mass and baryonic mass enclosed at that radius respectively. Nevertheless, the difference between the ``enclosed mass'' method and the ``direct summation'' method is negligible.
The RAR of this sample is in good agreement with the result obtained using the EAGLE simulation, and also has a small scatter around the median relation. Yet the RAR of this sample still differs from that of the SPARC sample.

In \autoref{T:tab1} we list all the values of the best-fit parameters for each sample with both the functional form introduced in \autoref{sec:rar}, and also the corresponding residuals. 
In many cases, the residual values in simulation do not closely follow a normal distribution; hence we report the median and the $16^\text{th}$ and the $84^\text{th}$ percentiles (in the notation of $m^{+1\sigma}_{-1\sigma}$). 
The error values of each of the fitting parameters reported in the table are the random (statistical) error, not the scatter in the distribution. Because of the large number of data points (roughly 100 locations for each galaxy in our sample), we obtain much smaller random error values. 

\section{Discussion and Summary}
\label{sec:discussion}

In this paper, we investigated the RAR in the disk galaxies of the MassiveBlack-II simulation at redshift $z=0.06$. A recent study by \cite{2016arXiv160905917M}, found a strong correlation between the observed radial acceleration and that due to baryons in the SPARC sample of galaxies for a wide range of luminosities, surface brightnesses, galaxy sizes and morphologies. This relation has been investigated in galaxy formation simulations by \cite{2016arXiv161006183K} in the MUGS2 cosmological zoom-in simulations and \cite{2016arXiv161007663L} in the EAGLE simulation. However, the samples of simulated galaxies in these studies span a relatively limited range of luminosities and surface densities. In contrast, the galaxies in our sample have been morphologically identified based on a dynamical bulge-disc decomposition and span a significantly wider range of luminosities and surface brightnesses than previous theoretical studies. 

In accord with the results of both observations (the SPARC sample) and other simulations, we find that the radial acceleration due to baryons is strongly correlated with the total radial acceleration at a variety of points throughout galactic disks and that this correlation exhibits fairly limited scatter. 
This result seems to imply that galaxy formation in a $\Lambda$CDM universe can, plausibly, induce a remarkably tight correlation between $\gobs$ and $\gbar$.
Previous work has argued that the dark matter haloes, which dominate the local gravitational potential, may lead to a tight relation of this kind simply because galaxy sizes do not vary significantly and because haloes are nearly self-similar \citep[e.g.,][]{kaplinghat_turner2002,navarro_etal2016}. 
Indeed, a large body of circumstantial evidence suggests that the properties of dark matter haloes play the critical role in determining the properties of galaxies that lie within them.

However, the RARs of our simulated galaxy samples are also clearly distinct from the observed RAR of the SPARC sample.
The slope of the RAR in the MB-II simulation is steeper than the asymptotic slope of the SPARC sample, and we do not see clear evidence for a transition from this asymptotic behavior in the acceleration range that we can probe. When being fitted to the SPARC functional form, \autoref{eq:gobsgbar}, the MB-II galaxies exhibit a significantly larger $\gdagger$ value than the SPARC result. The slight deviation from the power-law fit at the high-acceleration end in \autoref{F:fig_gobsgbar_all} does hint at a transition to the regime where $\gobs = \gbar$, yet in the regime where we have most data points there is not a clear transition.

The lack of this transition in our simulation is in part due to a very limited number of data points that have large acceleration, which reside in the inner region of massive galaxies. Although we do the best to mimic the SPARC population, the distribution of disk luminosities and surface brightnesses available in the MB-II simulation is still significantly different from that of the SPARC sample, particularly with a deficit of high surface brightness galaxies. Also, the finite gravitational softening length prevents accurate simulation of the high acceleration inner regions of massive disks.
A statistical sample of individual disks which span the transition region can provide one of the most incisive tests of the observed RAR and pin down the transition acceleration.
This test, however, would require simulations that have both better resolution and larger volumes, which are still beyond our current reach.

The effect of the finite resolution is twofold. First, resolution limits the number of particles in smaller galaxies and results in larger shot noise. Second, the finite spatial resolution limits the baryonic physics that can be resolved, which creates larger uncertainty (and possibly systematic biases). 
For example, the resolution limit manifests in our investigation of the bright and faint samples, as we find that the scatter (or residual with respect to the fitted relation) is significantly smaller in the bright sample. These two samples also have different median RARs. The best-fit RAR of the bright sample is very close to that of the full disk sample. When we remove data points that are within four times of the softening length ($4\epsilon$), the resulting RAR stays approximately unchanged, but with much smaller scatter at the low-acceleration range, since the best-fit RAR of the full disk sample is dominated by the bright sample.

In this work we also made a direct comparison with the result from the EAGLE simulation \citep{2016arXiv161007663L}, by selecting a similar sample of galaxies. The results from the MB-II and EAGLE simulations are in good agreement, and both have smaller scatter. It appears that for this sample of central galaxies, the difference in the detail baryonic feedback between the MB-II and EAGLE simulations does not affect the RAR significantly. This is an interesting result in and of itself, as it suggests that the RAR is a quantity that is robust to these choices. However, the RARs in both simulations differ from that of the observational measurements using the SPARC sample. The EAGLE and MB-II simulations also have similar mass resolution (while EAGLE adopts a smaller softening length), and hence both are subject to the impact of finite resolution as discussed above.

The degree to which the RAR is sensitive to baryonic physics implementations 
in different simulations remains somewhat unclear. 
For example, \citet{2016arXiv161006183K} found that the RAR evolves with redshift in the MUGS2 sample, and \citet{2016arXiv161007663L} find that the relation remains unchanged in the EAGLE sample. At face value this discrepancy seems to suggest that the evolution of the RAR could be sensitive to the implementation of baryonic physics. However, we should note that the 
MUGS2 and EAGLE samples span very different galaxy populations, as shown in \autoref{F:fig_lum}. Hence, it is possible that the aforementioned 
discrepancy is caused by the mass (or luminosity) dependence of the 
evolution of the RAR. 
The MUGS2 simulations, though with a small sample size and a limited luminosity range, have better resolution compared to the EAGLE simulation, which may also contribution to the different evolution of the RAR.
In this work we limit our study to the present-day population, and leave the redshift evolution of the RAR to follow-up studies. 

In conclusion, the RAR does pose a significant challenge for the understanding of galaxy formation, whether within the context of the standard $\Lambda$CDM scenario or in some other models. 
Any viable model of galaxy formation must yield the observed RAR. Therefore, the RAR represents a new and interesting test with which to confront simulations of galaxy formation in the hope of improving our understanding of the mechanisms that lead to the illumination of the universe.
Though current simulations do lead to an RAR with little scatter, in detail the simulated RAR differs significantly from that in the observed SPARC sample. Questions such as whether the RAR has a clear transition to $\gobs=\gbar$ and whether the RAR depends on galaxy luminosity and surface brightness also need to be answered. Further investigation of these issues will require larger-volume galaxy formation simulations and higher-resolution zoom-in simulations of individual galaxies.

\section*{Acknowledgements}
  We thank Stacy McGaugh and Jeffery Newman for helpful discussions.
  The MassiveBlack-II simulation was run on the Cray
XT5 supercomputer Kraken at the National Institute for Computational Sciences.
  YYM is supported by the Samuel P.\ Langley PITT PACC Postdoctoral Fellowship.
  RACC and TDM acknowledge support from
  NSF AST-1616168, NSF ACI-1614853,
NSF ACI-1036211, NSF AST-1517593, NSF AST-1009781,
NSF OCI-0749212, the NSF PAID program
  and RACC from NSF AST-1412966.
  AK and FZ have been supported by NSF AST-1312380. FZ acknowledges support from the Andrew Mellon Predoctoral Fellowship.
  ARZ has been supported by NSF AST-1517563 and NSF AST-1516266.

This research made use of many community-developed or community-maintained software packages, including (in alphabetical order):
Matplotlib \citep{matplotlib},
NumPy \citep{numpy},
and SciPy \citep{scipy}.
We also use the NASA Astrophysical Data System for bibliographic information.

\bibliographystyle{mnras}
\bibliography{references,software}

\bsp	
\label{lastpage}
\end{document}